# Reversible magnetization of MgB$_2$ single crystals with a two-gap nature

Byeongwon Kang,[a)] Heon-Jung Kim, Min-Seok Park, Kyung-Hee Kim, and Sung-Ik Lee[b)]

*National Creative Research Initiative Center for Superconductivity and Department of Physics,*

*Pohang University of Science and Technology, Pohang 790-784, Republic of Korea*

**Abstract**

We present reversible magnetization measurements on MgB$_2$ single crystals in magnetic fields up to 2.5 T applied parallel to the crystal's *c*-axis. This magnetization is analyzed in terms of the Hao-Clem model, and various superconducting parameters, such as the critical fields [ $H_c(0)$ and $H_{c2}(0)$ ], the characteristic lengths [ $\xi(0)$ and $\lambda(0)$ ], and the Ginzburg-Landau parameter, $\kappa$ are derived. The temperature dependence of the magnetic penetration depth, $\lambda(T)$, obtained from the Hao-Clem analysis could not be explained by theories assuming a single gap. Our data are well described by using a two-gap model.

# I. Introduction

MgB$_2$ with a superconducting transition temperature ($T_c$) of 39 K [1] has attracted great attention because it has several notable features compared to conventional superconductors. First of all, its $T_c$ of 39 K may be too high to be explained within the conventional electron-phonon mechanism. Therefore, an unconventional pairing mechanism [2] was proposed as a possible candidate of theoretical description. However, an early isotope experiment ruled out the unconventional theory and showed that the main driving force for the superconductivity is electron-phonon coupling [3,4]. Furthermore, it was suggested that the anisotropy in the electron-phonon coupling plays an important role in the unusually high $T_c$ [5-7].

Another notable feature of MgB$_2$ is its multi-gap property. A number of theoretical [5, 6, 8] and experimental investigations [9-15] suggest that MgB$_2$ has two different superconducting gaps: a larger gap originating from a two dimensional cylindrical Fermi surface with an average gap value of 6.8 meV and a smaller gap associated with a Fermi surface of three-dimensional tubular networks with an average gap value of 2.5 meV. Recently, direct evidence for two superconducting gaps was obtained from several measurements, such as specific heat [16], penetration depth [17, 18], tunneling [19, 20], point-contact spectroscopy [21] and photoemission spectroscopy [22, 23] on MgB$_2$ single crystals and thin films.

In addition, MgB$_2$ is a very interesting system regarding its vortex phases. Like high-$T_c$ superconductors, MgB$_2$ is reported to show various vortex phases [24], vortex phase transitions [25], and even peak effect [25-27] as a precursor of the vortex melting transition.

In this sense, MgB$_2$ may offer a unique opportunity to study the interplay between the various vortex phases and the two superconducting gaps.

To understand the vortex dynamics in MgB$_2$, accurate determination of material's parameters and their temperature dependence is necessary. Thermodynamic parameters such as the thermodynamic critical field and the magnetic penetration depth can be determined by the reversible magnetization study. Due to strong pinning, the measurements can not be carried out using thin films and polycrystalline samples [24]. On the other hand, the pinning is orders of magnitude lower in single crystals [28, 29] and therefore accurate determination of superconducting parameters can be carried out using single crystals. In an earlier study on single crystals, these parameters have been determined using the London model [30]. Since the London model only considers the free energy from electromagnetic contributions and ignores the free energy of core parts, which becomes quite important in low-$\kappa$ $(=\lambda/\xi)$ materials such as MgB$_2$, a description within the London model has inevitable limitations. In addition, since the upper critical fields of MgB$_2$ single crystals are quite small, the magnetic fields applied in that experiment are far above the range of the London model. Therefore, a more complete model, such as Hao-Clem's general model, which considers the free energies both from the electromagnetic part and from the core part is needed. The field range of this general model contains both low field London limit and high field Abrikosov limit.

In this paper, we present reversible magnetization measurements on MgB$_2$ single crystals with $T_c \cong 38$ K. The weak pinning property of our single crystals enabled us to have a wide reversible region in the magnetization data. To calculate more reliable superconducting parameters, we applied the Hao-Clem model [31] based on the Ginzburg-Landau (GL) theory to the magnetization. Using applied fields up to 2.5 T, we obtained superconducting parameters such as the critical field $H_c(0)$ and the coherence length $\xi(0)$. We also investigated the temperature dependence of the penetration depth, $\lambda(T)$, and found that our data are well described by using a two-gap model.

**II. Experiment**

Single crystals were grown using high pressures as described earlier [28, 29]. Briefly, a 1:1 mixture of Mg and amorphous B powders was well ground and pressed into a pellet. The pellet was put in a BN container and then placed in a high-pressure cell equipped with a graphite heater. The sample was heated to a temperature of ~ 1500 °C for 60 minutes inside a 14-mm cubic multi-anvil-type press (Rockland Research Corp.) under 3.5 GPa. After the heat treatment, the sample was slowly cooled to ~ 900 °C at a rate of 2 °C/min followed by a fast cool to room temperature.

Two sets of single crystals were investigated using magnetization measurements. In the first set, we collected 10 relatively hexagonal-shaped single crystals [28], with typical dimensions of 200 x 100 x 25 $\mu m^3$ on a substrate without an appreciable magnetic background and with their *c* axis aligned perpendicular to the substrate surface. The total volume of the collected single crystals was carefully calculated based on the images obtained using a polarizing optical microscope.

In the second set, we mounted a shiny and flat, but not hexagonal-shaped single crystal with dimensions of 800 x 300 x 60 $\mu m^3$ on a substrate. The values of $T_c$ and the transition width $\Delta T_c$ (10% - 90%) determined from the low-field magnetization data were 36.8 K and 2.5 K for the first set and 37.9 K and 1.4 K for the second set. Regardless of slightly different values of $T_c$, these two sets of crystals did not show any significant differences upon the magnetization analysis reported below. Therefore, in the following we discuss data obtained from the second set.

The measurement of the reversible magnetization was carried out by using a superconducting quantum interference device magnetometer (Quantum Design, MPMS-XL) with the field parallel to the *c*-axis of the sample.

## III. Results and Discussion

Figure 1 shows the zero-field-cooled (ZFC) and field-cooled (FC) magnetizations measured at a field of 10 G. The onset of superconducting transition is at 37.9 K with a transition width $\Delta T_c$ (10% - 90%) ~ 1.4 K. At $T = 5$ K, the value of $4\pi M/H$ for the ZFC is around 3 due to a demagnetization effect. The calculated demagnetization factor is about 0.67, and this large value is consistent with the plate-like shape of our sample. The inset of Fig. 1 shows the temperature dependence of the upper critical field $H_{c2}$ and the irreversibility field $H_{irr}$ of MgB$_2$. $H_{c2}(T)$ (In this paper, $H_{c2}$ refers to $H_{c2//c}$) was determined from the onset of superconductivity in the $4\pi M(T)$ curves obtained at different fields. A linear fit of $H_{c2}(T)$ near $T_c$ indicated a "bulk" $T_c$ of 37.1 K. $H_{c2}(0)$ was determined from the BCS-type function $H_{c2}(T) = H_{c2}(0)\left[1 - (T/T_c)^\alpha\right]^\beta$ [32], using the bulk $T_c$, with $H_{c2}(0)$, $\alpha$ and $\beta$ as fitting parameters. $H_{c2}(0)$ was found to be 2.80 T with $\alpha = 1.9$ and $\beta = 1.2$, which are in a reasonable range. The irreversible field $H_{irr}(T)$, where the ZFC and FC magnetizations start to diverge, was determined by using a simple criterion of $M_{FC}/M_{ZFC} = 0.95$ from the $4\pi M(T)$ curves obtained at different fields. $H_{irr}$ approaches $H_{c2}(0)$ below 10 K which results in a narrower reversible region at $T \leq 10$ K.

Figure 2 shows the temperature dependence of the reversible magnetization, $4\pi M(T)$, measured in the field range 0.1 T ≤ $H$ ≤ 2.5 T. Temperatures corresponding to the irreversibility line are indicated by arrows. The curves shift to lower temperatures as the field is increased and this feature is similar to that for conventional superconductors [33] and for infinite-layer superconductor $Sr_{0.9}La_{0.1}CuO_2$ [34]. Observed systematic shift of the magnetization is a typical mean field behavior in conventional superconductors, but is quite different from the high-$T_c$ superconductors. The thermal fluctuation effect [35] observed in most cuprate superconductors [36-38] is not significant in this system. The slope of magnetization, $d(4\pi M)/dT|_{T_c}$, is found to vary with the field and decreases by one order of magnitude as the magnetic field is increased from 0.1 T to 2.0 T, which is not expected from both the Abrikosov and London models. This is because the field range applied in this experiment covers from the low-field London limit to the high-field Abrikosov limit.

To analyze magnetization data obtained in a wide field range, the Hao-Clem model [31] was applied. By considering not only the electromagnetic energy outside of the vortex cores, but also the free energy changes arising from the cores, this variational model permits a reliable description of the reversible magnetization in the *entire* mixed state and an accurate determination of the thermodynamic parameters [32, 37, 39].

In the Hao-Clem model, the reversible magnetization in dimensionless form, $4\pi M' \equiv 4\pi M / \sqrt{2} H_c(T)$, is a universal function (temperature independent) of magnetic field, $H' \equiv H / \sqrt{2} H_c(T)$, for a given value of the GL parameter $\kappa$ [31]. Experimental $4\pi M$ versus $H$ data obtained at each temperature were fitted to the Hao-Clem model with $H_c(T)$ and $\kappa$ as parameters. If the value of $\kappa$ is appropriately chosen, the values of $H_c(T)$ should be the same for different fields, and the optimum value of $\kappa$ is obtained to give the smallest deviation of $H_c(T)$. Using this procedure $\kappa$ was found to be nearly temperature independent with an average value of 6.4 in the temperature range of 12 K $\leq T \leq$ 31 K. Using optimum values of $H_c(T)$ $4\pi M(H)$ data obtained at different temperatures were seen to collapse into a single curve when plotted as $4\pi M'$ vs. $H'$. Experimental data plotted in this manner is shown along with a theoretical Hao-Clem function corresponding to the average value of $\kappa$ in the inset of Fig. 3. It is obvious that our data covers a wide field region from the London limit where $H \ll H_{c2}$ to the Abrikosov limit where $H \approx H_{c2}$. The slight deviations from the universal curve at both ends of the data were caused by using the average $\kappa$.

Figure 3 shows the thermodynamic critical field $H_c$ versus temperature plot obtained from this analysis. The large errors of $H_c$ at low temperatures were caused by scattering of data due to stronger background contribution in higher magnetic fields and by taking the average $\kappa$. The solid line represents the fit of the temperature dependence of $H_c$ to $H_c(T) = H_c(0)\left[1-(T/T_c)^2\right]$ [40]. The result for $H_c(T)$ yields $H_c(0)$ = 0.23 T and $T_c$ = 37.0 K, which correspond to a slope of $dH_c/dT = -$ 0.012 T/K near $T_c$. By using the relation $H_{c2}(T) = \sqrt{2}\kappa H_c(T)$, we calculated the upper critical field slope as $(dH_{c2}/dT)_{T_c} = -$ 0.11 $\pm$ 0.01 T/K. Since $H_{c2}(T)$ for $H // c$ determined from the $4\pi M(H)$ data shows a nearly linear behavior near $T_c$, similar to the previous reports on single crystals [30, 41], we used the Werthamer-Helfand-Hohenberg (WHH) formula [42] to estimate $H_{c2}(0)$. In the WHH formula, $H_{c2}(0) = 0.5758(\kappa_1/\kappa)T_c\left|dH_{c2}/dT\right|_{T_c}$ and $\kappa_1/\kappa$ is 1.26 and 1.20 in the clean and the dirty limits, respectively. $H_{c2}(0)$ was calculated to be 2.86 ± 0.12 T in the clean limit, which in turn, the value of coherence length $\xi(0)$ became 10.7 ± 0.4 nm as deduced using the relation $\xi(0) = [\phi_0/2\pi H_{c2}(0)]^{1/2}$. The value of $H_{c2}(0)$ estimated from the Hao-Clem model was consistent with the value of $H_{c2}(0)$ obtained from the magnetization measurements $H_{c2}(T)$ (inset of Fig. 1), supporting the validity of the Hao-Clem approach. The little lower value of $H_{c2}(0)$ than those of previous reports [30, 43] may indicate that our crystals were relatively free of defects.

Employing the values of $H_c(T)$ and $\kappa$ obtained from the Hao-Clem model, we calculated the magnetic penetration depth, $\lambda(T)$, (in the following, $\lambda$ refers to $\lambda_{ab}$) using the relation $\lambda(T) = [\kappa\phi_0/2\sqrt{2}\pi H_c(T)]^{1/2}$, where $\phi_0$ is the flux quantum. The result is shown in Fig. 4. The temperature dependence of $\lambda$ has been controversial, and quadratic, linear, and exponential dependences have been reported [18, 44-48].

For a system like $MgB_2$ which is found to have two different gaps, the existence of two gaps should be reflected in $\lambda(T)$ in the following way; the large gap have a significant impact on $\lambda(T)$ at higher temperatures while the temperature dependence of $\lambda$ for $T \ll T_c$ would be dominated by the small gap. Therefore, we tried to apply the two-gap model [17] to describe our $\lambda(T)$. Here, the theoretical $\lambda(T)$ was calculated using

$$\lambda^{-2}(T)/\lambda^{-2}(0) = 1 - 2\left[c_1\int_{\Delta_s}^{\infty}\left(-\frac{\partial f}{\partial E}\right)D_s(E)dE + (1-c_1)\int_{\Delta_L}^{\infty}\left(-\frac{\partial f}{\partial E}\right)D_L(E)dE\right], \qquad (1)$$

where $c_1$ is a parameter which determines the contribution of the small gap, $\Delta_s$ is the small gap, $\Delta_L$ is the large gap, $f$ is the Fermi-Dirac distribution function, and $D_{S(L)}(E) = E / [E^2 - \Delta_{S(L)}^2]^{1/2}$. Each parameter was allowed to vary only within a certain range determined from the earlier results [17, 18, 20]. The two-gap model using Eq. (1) describes our $\lambda(T)$ relatively well over the whole temperature region as plotted as a solid line. From this, we obtained the gap values $\Delta_s$ = 1.9 meV and $\Delta_L$ = 6.1 meV with relative proportion 4:6 and these gap values are in agreement with values obtained by other experiments [15, 17, 18, 20, 22]. The contribution of small gap is manifested by a plateau at low temperatures and nearly the same relative proportions of two gaps were reported on polycrystalline and single crystals [18]. From the two-gap model, $\lambda(0)$ of 76.4 nm was obtained.

The inset of Fig. 4 shows the temperature dependence of $\lambda^2(0)/\lambda^2(T)$, which represents the normalized superfluid density of MgB$_2$. As expected from Fig. 4, a good agreement with the two-gap model is achieved. A little plateau and then a downward curvature up to ~ 0.5 $T/T_c$ reflect higher contribution of small gap to the superfluid density. For comparison, the two-fluid model $\lambda^2(0)/\lambda^2(T) = 1 - (T/T_c)^4$ and the BCS predictions are also depicted. Even though a direct comparison with the theoretical models is not possible at low temperatures due to lack of data below 0.3 $T/T_c$, our data show obvious discrepancies from the two-fluid model and a single-gap BCS theory. All the parameters determined in this study are summarized in Table 1.

## IV. Summary

The reversible magnetization of MgB$_2$ single crystals was measured for magnetic fields up to 2.5 T. The reversible magnetization was analyzed using the Hao-Clem model.

Various superconducting parameters derived from this analysis are summarized in Table 1. The temperature dependence of $\lambda$ determined using Hao-Clem approach was found to be well described by a two-gap model with gap values $\Delta_s = 1.9$ meV and $\Delta_L = 6.1$ meV.

## ACKNOWLEDGMENTS

We thank P. Chowdhury and Mun-Seog Kim for many helpful discussions. This work was supported by the Ministry of Science and Technology of Korea through the Creative Research Initiative Program.

**References**

1. J. Nagamatsu, N. Nakagawa, T. Muranaka, Y. Zenitani, and J. Akimitsu, Nature (London) **410**, 63 (2001).

2. J.E. Hirsh, Phys. Lett. A **282**, 393 (2001)

3. S.L. Bud'ko, G. Lapertot, C. Petrovic, C.E. Cunningham, N. Anderson, and P.C. Canfield, Phys. Rev. Lett. **86**, 1877 (2001).

4. D.G. Hinks, H. Claus and J.D. Jorgensen, Nature **411**, 457 (2001).

5. A.Y. Liu, I.I. Mazin, and J. Kortus, Phys. Rev. Lett. **87**, 087005 (2001).

6. H.J. Choi, D. Roundy, H. Sun, M.L. Cohen, and S.G. Louie, Nature **418**, 758 (2002).

7. A. Shukla, M. Calandra, M. d'Astuto, M. Lazzeri, F. Mauri, C. Bellin, M. Krisch, J. Karpinski, S.M. Kazakov, J. Jun, D. Daghero, and K. Parlinski, Phys. Rev. Lett. **90**, 095506 (2003).

8. H.J. Choi, D. Roundy, H. Sun, M.L. Cohen, and S.G. Louie, Phys. Rev. B **66**, 020513(R) (2002).

9. H. Schmidt, J.F. Zasadzinski, K.E. Gray, and D.G. Hinks, Phys. Rev. B **63**, 220504 (2001).

10. F. Giubileo, D. Roditchev, W. Sacks, R. Lamy, D.X. Thanh, J. Klein, S. Miraglia, D. Fruchart, J. Marcus, and P. Monod, Phys. Rev. Lett. **87**, 177008 (2001)

11. P. Szabó, P. Samuely, J. Kačmarčik, T. Klein, J. Marcus, D. Fruchart, S. Miraglia, C. Marcenat, and A.G.M. Jansen, Phys. Rev. Lett. **87**, 137005 (2001).

12. P. Seneor, C.-T. Chen, N.-C. Yeh, R.P. Vasquez, L.D. Bell, C.U. Jung, M.-S. Park, H.-J. Kim, W.N. Kang, and S.-I. Lee, Phys. Rev. B **65**, 012505 (2001).

13. H.D. Yang, J.-Y. Lin, H.H. Li, F.H. Hsu, C.J. Liu, S.-C. Li, R.-C. Yu, and C.-Q. Jin, Phys. Rev. Lett. **87**, 167003 (2001).

14. H. Kotegawa, K. Ishida, Y. Kitaoka, T. Muranaka, and J. Akimitsu, Phys. Rev. Lett. **87**, 127001 (2001).

15. X.K. Chen, M.J. Konstantinović, J.C. Irwin, D.D. Lawrie, and J.P. Franck, Phys. Rev. Lett. **87**, 157002 (2001).

16. F. Bouquet, Y. Wang, I. Sheikin, T. Plackowski, A. Junod, S. Lee, and S. Tajima, Phys. Rev. Lett. **89**, 257001 (2002).

17. M.S. Kim, J.A. Skinta, T.R. Lemberger, W.N. Kang, H.-J. Kim, E.-M. Choi, and S.-I. Lee, Phys. Rev. B **66**, 064511 (2002).

18. F. Manzano, A. Carrington, N. E. Hussey, S. Lee, A. Yamamoto, and S. Tajima, Phys. Rev. Lett. **88**, 047002 (2002).

19. M. Iavarone, G. Karapetrov, A.E. Koshelev, W.K. Kwok, G.W. Crabtree, D.G. Hinks, W.N. Kang, E.-M. Choi, H.J. Kim, H.-J. Kim, and S.-I. Lee, Phys. Rev. Lett, **89**, 187002 (2002).

20. M.R. Eskildsen, M. Kugler, S. Tanaka, J. Jun, S.M. Kazakov, J. Karpinski, and O. Fischer, Phys. Rev. Lett. **89**, 187003 (2002).

21. R.S. Gonnelli, D. Daghero, G.A. Ummarino, V.A. Stepanov, J. Jun, S.M. Kazakov, and J. Karpinski, Phys. Rev. Lett, **89**, 247004 (2002).

22. S. Tsuda, T. Yokoya, T. Kiss, Y. Takano, K. Togano, H. Kito, H. Ihara, and S. Shin, Phys. Rev. Lett. **87**, 177006 (2001).

23. S. Souma, Y. Machida, T. Sato, T. Takahashi, H. Matsui, S.-C. Wang, H. Ding, A. Kaminski, J. C. Campuzano, S. Sasaki, and K. Kadowaki, Nature **423**, 65 (2003).

24. Hyeong-Jin Kim, W. N. Kang, Eun-Mi Choi, Mun-Seog Kim, Kijoon H.P. Kim, Sung-Ik Lee, Phys. Rev. Lett. **87**, 087002 (2001).

25. M. Angst, R. Puzniak, A. Wisniewski, J. Jun, S. M. Kazakov, and J. Karpinski, Phys. Rev. B **67**, 012502 (2003).

26. U. Welp, A. Rydh, G. Karapetrov, W.K. Kwok, G. W. Crabtree, Ch. Marcenat, L. Paulius, T. Klein, J. Marcus, K.H.P. Kim, C.U. Jung, H.-S. Lee, B. Kang, and S.-I. Lee, Phys. Rev. B **67**, 012505 (2002).

27. M. Pissas, S. Lee, A. Yamamoto, and S. Tajima, Phys. Rev. Lett. 89, 097002 (2002).

28. C.U. Jung, J.Y. Kim, P. Chowdhury, Kijoon H.P. Kim, Sung-Ik Lee, D.S. Koh, N. Tamura, W.A. Caldwell, and J.R. Patel, Phys. Rev. B **66**, 184519 (2002).

29. Kijoon H.P. Kim, Jae-Hyuk Choi, C.U. Jung, P. Chowdhury, Hyun-Sook Lee, Min-Seok Park, Heon-Jung Kim, J.Y. Kim, Zhonglian Du, Eun-Mi Choi, Mun-Seog Kim, W.N. Kang, Sung-Ik Lee, Gun Yong Sung, and Jeong Yong Lee, Phys. Rev. B **65**, 100510 (2002).

30. M. Zehetmayer, M. Eisterer, J. Jun, S.M. Kazakov, J. Karpinski, A. Wisniewski, and H.W. Weber, Phys. Rev. B **66**, 052505 (2002).

31. Z. Hao and J.R. Clem, M. W. McElfresh, L. Civale, A. P. Malozemoff, and F. Holtzberg, Phys. Rev. B. **43**, 2844 (1991).

32. E. Helfand, and N.R. Werthamer, Phys. Rev. **147**, 288 (1966).

33. M. Suenaga, A.K. Ghosh, Y. Xu, and D.O. Welch, Phys. Rev. Lett. **66**, 1777 (1991).

34. Mun-Seog Kim, Thomas R. Lemberger, C.U. Jung, Jae-Hyuk Choi, J.Y. Kim, Heon-Jung Kim, and Sung-Ik Lee, Phys. Rev. B. **66**, 214509 (2002).

35. L.N. Bulaevskii, M. Ledvij, and V.G. Kogan, Phys. Rev. Lett. **68**, 3773 (1992).

36. Mun-Seog Kim, C.U. Jung, Sung-Ik Lee, and A. Iyo, Phys. Rev. B. **63**, 134513 (2001).

37. Mun-Seog Kim, Sung-Ik Lee, A. Iyo, K. Tokiwa, M. Tokumoto, and H. Ihara, Phys. Rev. B **57**, 8667 (1998).

38. Heon-Jung Kim, P. Chowdhury, In-Sun Jo, and Sung-Ik Lee, Phys. Rev. B. **66**, 134508 (2002).

39. Mun-Seog Kim, Sung-Ik Lee, Seong-Cho Yu, Irina Kuzemskaya, Efim S. Itskevich, K.A. Lokshin, Phys. Rev. B **57**, 6121 (1998).

40. C.J. Gorter and H.B.G. Casimir, Physica (Amsterdam) **1**, 306 (1934).

41. U. Welp, A. Rydh, G. Karapetrov, W.K. Kwok, G.W. Crabtree, C. Marcenat, L.M. Paulius, T. Klein, J. Marcus, K.H.P. Kim, C.U. Jung, H.-S. Lee, B. Kang, and S.-I. Lee, Physica C, **387**, 137-142 (2003).

42. N.R. Werthamer, E. Helfand, and P.C. Hohenberg, Phys. Rev. **147**, 295 (1966).

43. M. Angst, R. Puzniak, A. Wisniewski, J. Jun, S.M. Kazakov, J. Karpinski, J. Roos, and H. Keller, Phys. Rev. Lett. **88**, 167004 (2002).

44. A.K. Pradhan, M. Tokunaga, Z.X. Shi, Y. Takano, K. Togano, H. Kito, H. Ihara, and T. Tamegai, Phys. Rev. B **65**, 144513 (2002).

45. A.V. Pronin, A. Pimenov, A. Loidl, and S.I. Krasnosvobodtsev, Phys. Rev. Lett. **87**, 097003 (2001).

46. R.A. Kaindl, M.A. Carnahan, J. Orenstein, D.S. Chemla, H.M. Christen, H.Y. Zhai, M. Paranthaman, and D.H. Lowndes, Phys. Rev. Lett. **88**, 027003 (2002).

47. S.L. Li, H.H. Wen, Z.W. Zhao, Y.M. Ni, Z.A. Ren, G.C. Che, H.P. Yang, Z.Y. Liu, and Z.X. Zhao, Phys. Rev. B **64**, 094522 (2001).

48. R. Prozorov, R.W. Giannetta, S.L. Bud'ko, and P.C. Canfield, Phys. Rev. B **64**, 180501 (R) (2002).

Table 1. Transition temperature $T_c$, the GL parameter $\kappa = \lambda/\xi$, the thermodynamic critical field $H_c(0)$, the upper critical field $H_{c2}(0)$, the coherence length $\xi(0)$, and the penetration depth $\lambda(0)$ of MgB$_2$ derived from the reversible magnetization.

| $T_c$ (K) | $\kappa$ | $dH_{c2}/dT\|_{T_c}$ (T/K) | $H_c(0)$ (T) | $H_{c2}(0)$ (T) | $\xi(0)$ (nm) | $\lambda(0)$ (nm) |
|---|---|---|---|---|---|---|
| 37.9[a] 37.1[b] | 6.4 ± 0.6 | −0.10 ± 0.01 | 0.23 | 2.86 ± 0.12[c] 2.80[d] | 10.7 ± 0.4[c] 10.8[d] | 76.4[e] |

[a] from low field magnetization
[b] bulk $T_c$: from a linear fit of $H_{c2}(T)$ near $T_c$
[c] assuming the BCS clean limit
[d] from $H_{c2}(T)$
[e] from the two-gap model

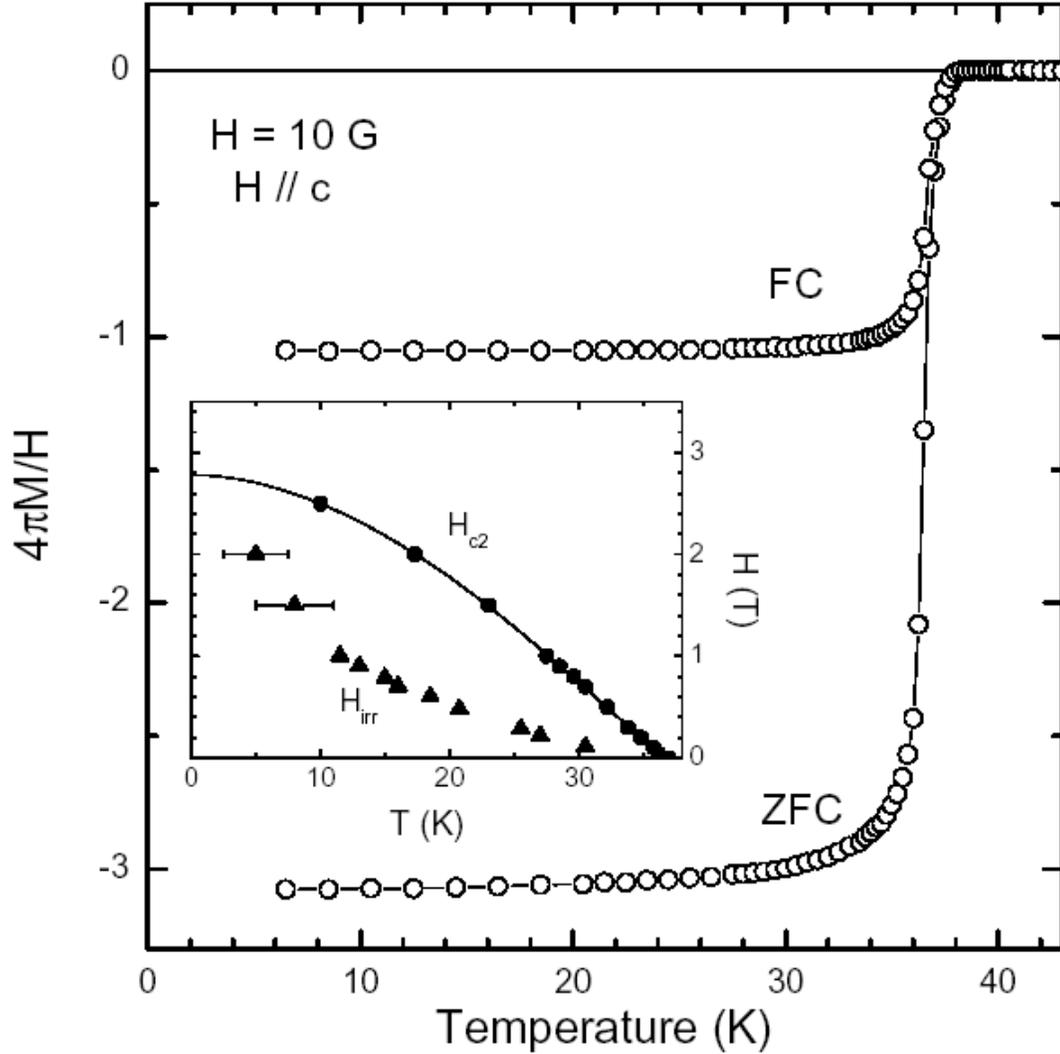

Fig.1. Temperature dependence of low-field magnetization, $4\pi M/H$, of a MgB$_2$ single crystal for $H = 10$ G. $T_c = 37.9$ K and $\Delta T_c$ (10%-90%) ~ 1.4 K. Inset: temperature dependence of $H_{c2}$ and $H_{irr}$ determined from the magnetization measurement. The solid line is a BCS-type function with $H_{c2}(0) = 2.80$ T.

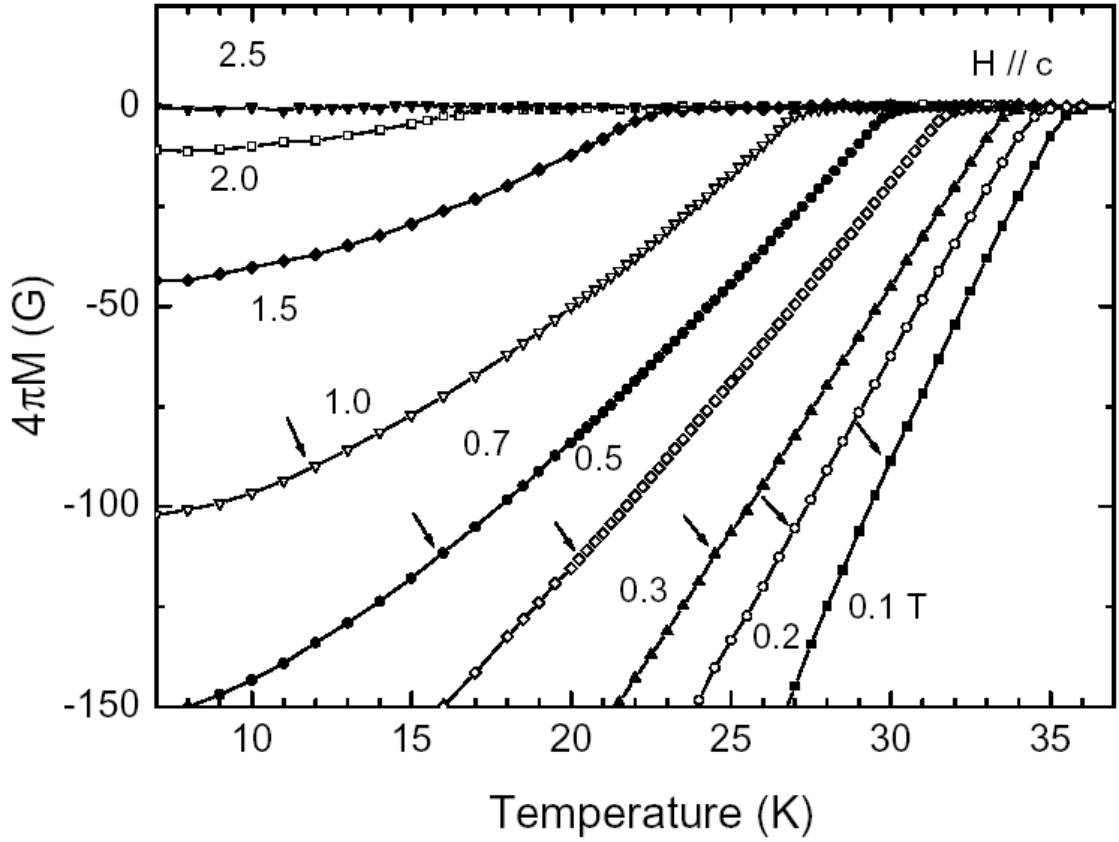

Fig. 2. Temperature dependence of the reversible magnetization, $4\pi M(T)$, in the field range $0.1\,\text{T} \leq H \leq 2.5\,\text{T}$. The irreversible temperatures are indicated as arrows.

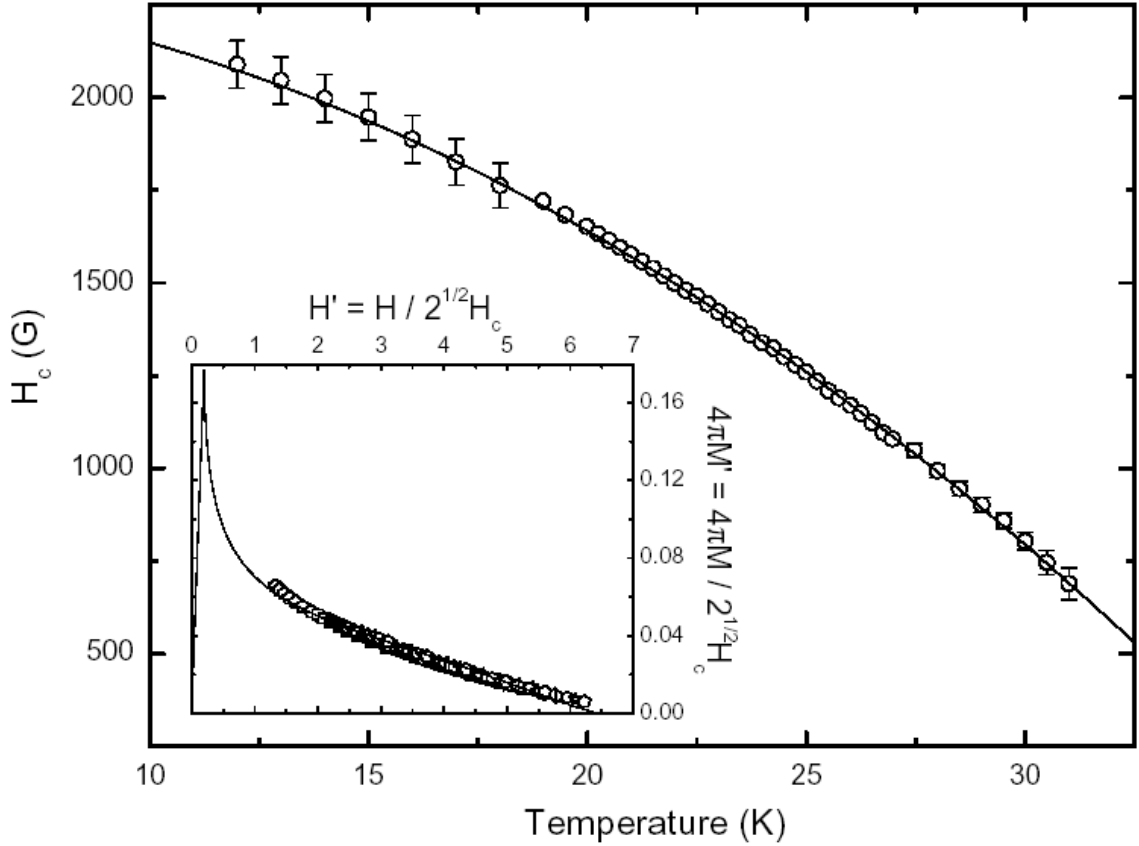

Fig. 3. Temperature dependence of the thermodynamic critical field, $H_c(T)$. The solid line represents $H_c(T) = H_c(0)\left[1-(T/T_c)^2\right]$. Inset: Magnetization, $-4\pi M' \equiv -4\pi M/\sqrt{2}H_c$ vs. external magnetic field, $H' \equiv H/\sqrt{2}H_c$. The solid line depicts the universal curve derived from the Hao-Clem model using $\kappa = 6.4$.

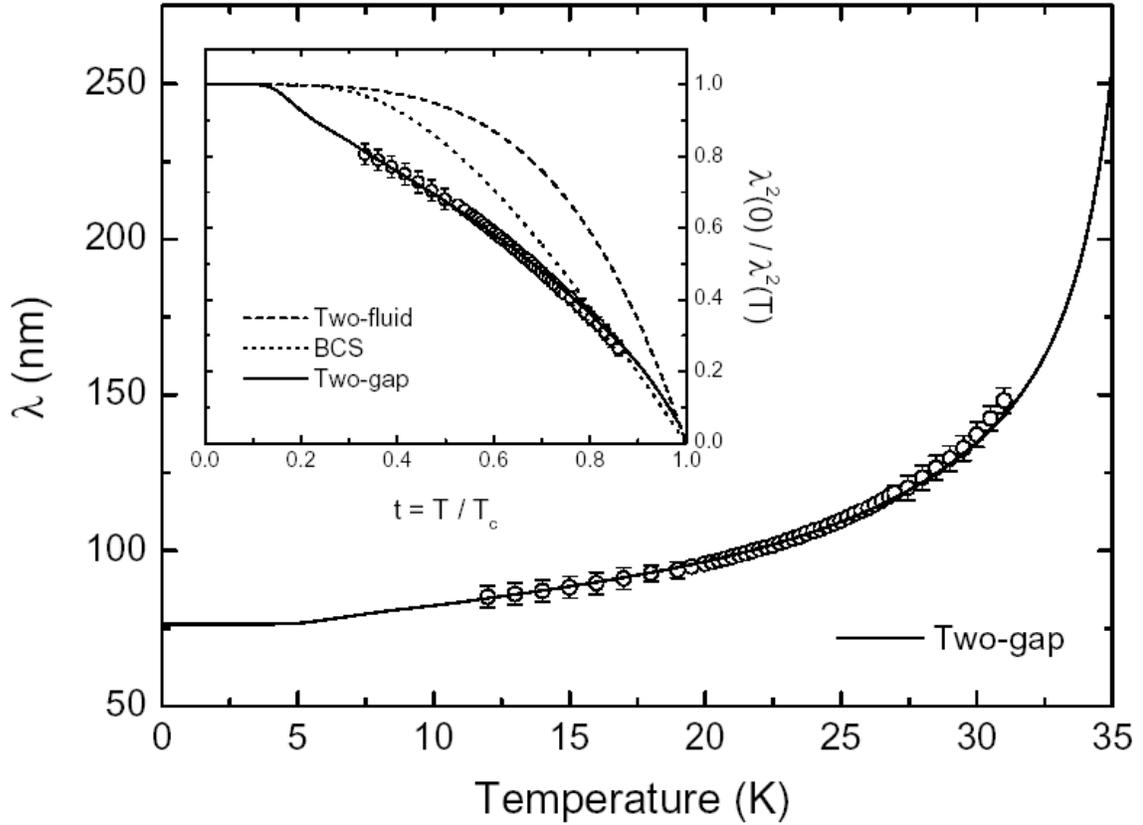

Fig. 4. Temperature dependence of $\lambda$ calculated from the Hao-Clem model. A theoretical calculation with the two-gap model is shown as a solid line. The formula of the curve is given in the text. Inset: temperature dependence of $\lambda^2(0)/\lambda^2(T)$ calculated from the Hao-Clem model. The solid line represents the two-gap model. The theoretical curves by the two-fluid model (dashed) and the BCS model (dotted) are also drawn.